\documentclass[twocolumn,showpacs,preprintnumbers,nofootinbib]{revtex4}
\usepackage{amsfonts}
\usepackage{graphicx}

\usepackage{graphicx}
\usepackage{dcolumn}
\usepackage{amssymb}
\usepackage{amsmath}
\usepackage{epsfig}

\usepackage[colorlinks=true,linkcolor=red,citecolor=blue,urlcolor=cyan,
bookmarks=true,bookmarksopen=true,pdfpagemode=None,pdfstartview=FitH]{hyperref}

\newcommand{\be}{\begin{equation}}
\newcommand{\ee}{\end{equation}}
\newcommand{\bea}{\begin{eqnarray}}
\newcommand{\eea}{\end{eqnarray}}
\def\lapp{\mathrel{\rlap{\raise.5ex\hbox{$<$}}
                    {\lower.5ex\hbox{$\sim$}}}}
\def\gapp{\mathrel{\rlap{\raise.5ex\hbox{$>$}}
                    {\lower.5ex\hbox{$\sim$}}}}

\begin{document}

\title { 
Long-range forces : atmospheric neutrino oscillation at a magnetized detector
}

\author
{\sf Abhijit Samanta\footnote{E-mail address: abhijit.samanta@gmail.com}}
\affiliation
{
{\em Ramakrishna Mission Vivekananda University, Belur Math, Howrah 711 202, 
India}\\
and\\ 
\footnote{past address where the work was initiated.}{\em Harish-Chandra Research Institute,
Chhatnag Road,
Jhusi,
Allahabad 211 019,
India\\ 
}}
\date{\today}

\begin{abstract}
\noindent
Among the combinations $L_e-L_\mu$,  $L_e-L_\tau$ and  $L_\mu-L_\tau$  
any one can be gauged in anomaly free way with the standard model gauge group. The masses of these
gauge bosons can be so light that it can induce long-range forces on the Earth due to the
electrons in the Sun. This type of forces can be constrained significantly
from neutrino oscillation. As the sign of the potential is opposite  for neutrinos and
 antineutrinos, a magnetized iron calorimeter detector (ICAL) would be able to 
produce strong constraint on it. We have 
made conservative studies of these long-range forces with atmospheric neutrinos at ICAL 
considering only the muons of charge current interactions.
We find stringent bounds on the couplings $ \alpha_{e\mu, e\tau} \lapp 1.65 \times
10^{-53}$ at 3$\sigma$ CL with an exposure of 1 Mton$\cdot$yr if there is no such force. 
For nonzero input values of the couplings we find that
the potential $V_{e\mu}$ opposes and $V_{e\tau}$ helps  
to discriminate the mass hierarchy.
However, both potentials help significantly 
to discriminate the octant of $\theta_{23}$. 
The explanation of the anomaly in recent MINOS data (the difference of $\Delta m^2_{32}$ 
for neutrinos and 
antineutrinos), using long-range force originated from the mixing of the gauge boson $Z^\prime$ 
of  $L_\mu-L_\tau$  with the standard model gauge boson $Z$,
can be tested at ICAL at more than 5$\sigma$ CL. 
We have also discussed how to disentangle this   from  the solution with CPT violation
using the seasonal change  of  the distance between the Earth and the Sun.
\end{abstract}


\maketitle
\section{Introduction}
The  large hadron collider at CERN will probe the extensions of standard model
above the electroweak scale. On the other hand, some of the extensions 
below the electroweak scale can be probed at the neutrino oscillation experiments. 
The extensions below the electroweak scale introduce massless or nearly massless gauge \cite{foot,axion} 
or Higgs bosons \cite{singletm,tripletm,hall,Chang:1989xm} and 
they couple with matter
very feebly and remain invisible. These lead to the existence of new kind of forces 
either i) generating deviations of gravitational law at short distances, or ii) predicting  low mass
particles whose exchange will induce forces at long distances, generally violating the equivalence 
principle \cite{Damour:1996xt,fischbach}.
A number of experiments have been searching for these 
new  forces. The null results provide bounds
on particle physics models, gravitational physics, and even on cosmological models
\cite{fischbach,Adelberger:2003zx,Williams:1995nq}.

The bounds on these couplings to  baryon and/or lepton number \cite{Lee:1955vk}  can be obtained 
from the testing  of equivalence principle \cite{Eotvos:1922pb} 
(the free fall acceleration is same for all bodies independent of their chemical content). 
In \cite{Okun:1995dn}, the author has used this idea to establish a bound on
the strength of an hypothetical vectorial leptonic force and
 obtained the bound on the ``fine structure'' constant
$\alpha \lapp 10^{-49}.$
See \cite{Dolgov:1999gk} for a review.  
A comparable limit also comes from 
lunar laser ranging \cite{Williams:1995nq}, which measures the differential 
acceleration of the Earth and the Moon towards the Sun.

However, one can extend the SM with an additional $U(1)$ gauge symmetry 
 without introducing any anomaly  
for one of the 
lepton flavor combinations: $L_e-L_\mu$, $L_\mu-L_\tau$, and $L_e-L_\tau$.
The  masses of the gauge bosons can be so light that the induced
forces may have terrestrial range. Then the 
electrons inside the Sun can induce forces on the Earth surface depending on the lightness 
of the gauge boson.  
These forces on the Earth may also be from supernova neutrinos, 
galactic electrons  depending on its ranges. 
These couple
only to electron (and neutrino) density inside a massive object.
As a result, the acceleration experienced by an object depends on its leptonic content and mass;
and thus violates equivalence principle. 

The long-range (LR) forces can play role in neutrino oscillation. 
The relatively stronger bounds than those from testing of the equivalence principle 
 have been obtained from solar, atmospheric and supernova data 
\cite{Joshipura:2003jh, Bandyopadhyay:2006uh, GonzalezGarcia:2006vp, Grifols:1993rs, 
Grifols:2003gy}\footnote{There are plenty relic neutrinos and antineutrinos in the universe 
($\sim 50/cm^3$),
which may screen the leptonic charges of the of celestial bodies. However, it has been analyzed 
in detail and shown that the screening is impossible \cite{Blinnikov:1995kp}}.
The long-range forces due to galactic electrons, which can affect the galactic rotation curves 
if the range $R_{LR} \gapp R_{\rm gal}$ (where $R_{\rm gal}$ is the distance from the galactic center $\sim$ 10 kpc),
is also now very tightly constrained \cite{Bandyopadhyay:2006uh}. Finally, all these
results are consistent with the existing bounds on violation of equivalence principle.

The sign of this potential is opposite for neutrinos and  anti-neutrinos
and hence can lead to apparent differences in neutrino and anti-neutrino oscillation
probabilities without introducing CP or CPT violation. For instance, the recently
found discrepancy in the survival probabilities of $\nu_\mu$s and $\bar\nu_\mu$s 
in the MINOS experiment \cite{pvahle}  has been explained using the mixing of $Z^\prime$
boson of
$L_\mu-L_\tau$ symmetry
with the $Z$ boson  of the SM model;
  and the required associated parameters to explain this anomaly can be
found in \cite{Heeck:2010pg}.

We have studied the long-range forces with atmospheric neutrinos at the magnetized iron 
calorimeter detector (ICAL)  proposed at the India-based neutrino Observatory (INO) 
\cite{Arumugam:2005nt}, which can directly measure the potential
detecting  separately $\nu_\mu$s and $\bar\nu_\mu$s.
In this article we have  focussed on:

- the bounds on the couplings of long-range forces, 

- how significantly long-range forces modify the sensitivity
of the measurement of  oscillation parameters; particularly,
the octant of 2-3 mixing (sign of $\delta^{\rm oct}=\theta_{23}-45^\circ$) 
and the mass hierarchy (sign of $\Delta m_{31}^2$),

- test of the explanation of the MINOS data with long-range forces \cite{Heeck:2010pg}.

- possibility to disentangle the effect of CPT and long-range forces
\footnote{The advantages of atmospheric neutrinos to discriminate 
$CPT$ violation from $CP$ violation and
nonstandard interactions have been  discussed in \cite{Samanta:2010ce}.}.

We have also studied the changes of  sensitivity to mass hierarchy 
and to $\theta_{23}$-octant with true (input) $\theta_{23}$ values considering no such 
forces (This was not studied in earlier works.).

The paper is organized as follows. We discuss
the full three flavor neutrino oscillation in matter
with long-range potential in sec. \ref{s:osc}, the details of the
analysis method in sec. \ref{s:chi}, and
the bounds 
on the couplings  in Sec. \ref{s:bounds}.  The effects of the long-range potential on discrimination of mass hierarchy 
and on $\theta_{23}$-octant is discussed in sec. \ref{s:23sector}.  Finally, the  testing of the explanation of the 
anomaly in MINOS data
using long-range force is described in Sec. \ref{s:minos}. 
The discussion and conclusion are given in \ref{s:dis}.



%
%
%

\section{Oscillation in presence of long-range potential}\label{s:osc}


The electrons inside the Sun generate a potential at the Earth by \cite{Joshipura:2003jh}:
\be V_{e\mu,e\tau} = \alpha_{e\mu,e\tau}\frac{N_e}{R_{ES}}\approx 1.3\times 10^{-11}{\rm eV} 
\left( \alpha_{e\mu,e\tau}/10^{-50}\right),\ee 
where, $\alpha_{e\mu,e\tau}=g^2_{e\mu,e\tau}/{4\pi}$;  $g_{e\mu,e\tau}$ are the gauge 
couplings of $L_e-L_\mu$ and  $L_e-L_\tau$ symmetries. $N_e\approx 10^{57}$ is the number of electrons
inside the Sun \cite{sun}  and $R_{ES}=7.6\times 10^{26}$ GeV$^{-1}$, the distance between the 
Earth and the Sun.

In a three neutrino framework the neutrino flavor states $|\nu_\alpha
\rangle$, $\alpha = e, \mu, \tau$ can be expressed as  linear superpositions
of the neutrino mass
eigenstates $|\nu_i\rangle$, $i=1,2,3$ with
masses $m_i$ :
\begin{equation}
\vert\nu_\alpha \rangle = \sum_i U_{\alpha i} \vert\nu_i\rangle~.
\end{equation}
$U$ is the  $3 \times 3$ unitary matrix. 
The time evolution of the flavor states is
\begin{equation}
i\frac{\rm d}{\rm d t}\left[ \nu_\alpha \right] =\frac{1}{2E}
U M_\nu^2 U^\dagger
\left[\nu_\alpha\right],
\end{equation}
where, $[\nu_{\alpha}]$ is the vector of flavor eigenstates, and
$\left[\nu_{\alpha}\right]^T =
\left[\vert\nu_e\rangle,\vert\nu_{\mu}\rangle,\vert\nu_{\tau}\rangle \right]$.

The evolution equation in the presence of matter and long-range potential is 
\begin{eqnarray}
i\frac{\rm d}{{\rm d} t} \left[ \nu_\alpha \right] = \frac{1}{2E}
\left[U M_{\nu}^2 U^{\dagger}
 +\left(
          \begin{array}{ccc}
          A & 0 & 0  \\
          0 & 0 & 0  \\
          0 & 0 & 0 \end{array}\right)
+V_{LR}
\right]
\left[ \nu_\alpha \right],
\label{eq:ham}
\end{eqnarray}
where,
\begin{eqnarray}
V_{LR}=
\left(
          \begin{array}{ccc}
          V_{e\mu} & 0 & 0  \\
          0 & -V_{e\mu} & 0  \\
          0 & 0 & 0 \end{array}\right)
{\rm or }\left(
          \begin{array}{ccc}
          V_{e\tau} & 0 & 0  \\
          0 & 0 & 0  \\
          0 & 0 & -V_{e\tau} \end{array}\right).
\end{eqnarray}

The matter term 
$
A = 2 \sqrt{2} G_F n_e E = 7.63 \times 10^{-5}~{\rm eV}^2~\rho({\rm gm/cc})~
E({\rm GeV})~ \hbox{eV}^2.
$
\label{densm}
%
Here, $G_F$, $n_e$ and $\rho$ are the Fermi constant, the electron number density
and the matter density of the medium, respectively. The evolution equation for
antineutrinos has the reversed sign for $A$, $V_{LR}$ and the phase
$\delta$.
We have numerically solved  full three flavor oscillation in presence
of matter and long-range forces.

However, to understand the bounds it is easy to consider  two flavor 
$\mu-\tau$ oscillation, (which has been  studied in  \cite{Joshipura:2003jh} 
to constrain the long-range forces using  atmospheric neutrino data of Super-Kamiokande experiment 
\cite{Super-K}). 
To understand the effect of 1-3 mixing, one needs to consider the changes of the 
effective oscillation parameters
due to $V_{LR}$.
The $\mu-\tau$ oscillation in presence of $V_{e\tau}$ is governed by the evolution equation
\begin{eqnarray}\label{H}
&& i \frac{d}{dt}  \begin{pmatrix}
 \nu_\mu  \\
   \nu_\tau \
  \end{pmatrix}= \nonumber \\
&& \begin{pmatrix}
    -\frac{\Delta m_{32}^2}{4 E} \cos 2 \theta_{23} & \frac{\Delta m_{32}^2}{4 E} \sin 2 \theta_{23}  \\
    \frac{\Delta m_{32}^2}{4 E} \sin 2 \theta_{23}  & \frac{\Delta m_{32}^2}{4 E} \cos 2 \theta_{23}  - V_{e \tau} \
  \end{pmatrix}\begin{pmatrix}
    \nu_\mu  \\
    \nu_\tau \
  \end{pmatrix}
\end{eqnarray}
Then, the survival probability of $\nu_\mu$   
\begin{equation}\label{p}
  P_{\mu \mu}= 1- \sin^2 2\tilde\theta_{23} ~\sin^2 \frac{\Delta\tilde  m_{23}^2
  L}{4 E},
\end{equation}
where, $L$  is the neutrino flight path length. The effective mixing angle $\tilde\theta_{23}$
and  $\Delta\tilde m_{32}^2$ are related with their vacuum quantities by the relations 
\begin{equation}\label{e:sin}
  \sin^2 2 \tilde\theta_{23} = \frac{Sin^2 2 \theta_{23}}
{\left[(\xi_{e \tau}-\cos 2 \theta_{23})^2 +
  \sin^2 2 \theta_{23}\right]}
\end{equation}
and
\begin{equation}\label{e:dM}
  \Delta \tilde m_{23}^2 =\Delta m_{23}^2 \left[(\xi_{e \tau}-\cos 2
\theta_{23})^2 +
  \sin^2  2\theta_{23})^{1/2}\right];
\end{equation}
where,
 $ \xi_{e \tau} \equiv  \frac{2 V_{e \tau} E}{\Delta m_{32}^2}$.
%
The potential $V_{e \tau}$ and the corresponding $\xi$ change sign for $\bar\nu$.
Similarly, for $L_e-L_\mu$ gauge symmetry the survival probability can be obtained
and they satisfy 
\begin{equation}\label{e:ppba}
  P_{\mu \mu}(V_{e \tau})= P_{\bar \mu \bar \mu}(-V_{e \tau})=P_{\mu \mu}(-V_{e
  \mu}) = P_{\bar \mu \bar \mu}(V_{e \mu})
\end{equation}

\section{The $\chi^2$ analysis}\label{s:chi}
To evaluate the potential of ICAL with atmospheric neutrinos  
we generate events by NUANCE-v3 \cite{Casper:2002sd} and
consider only the muon energy and its direction (directly measurable quantities)
of the events for a conservative estimation. 
The energy and
angular resolutions of the muons  at ICAL are very
high: 4-10\% for energy and 4-12\% for zenith angle,  which are obtained from  GEANT \cite{geant} 
simulation. The ranges are due to different energies and different angles with respect to the
iron layers. These uncertainties are very negligible compared to the uncertainties
in reconstructed neutrinos due to kinematics of the scattering processes.  

The major uncertainty arises from the particles produced in the event other than muon. 
One might expect that consideration of the hadrons for neutrino energy $\gapp 2$ GeV will 
improve the sensitivities substantially.
But, in \cite{Samanta:2010xm}, it has been found that there is a very marginal improvement on 
measurement of $\theta_{23}$ and a small improvement on $\Delta m_{31}^2$:
$\delta (\Delta m_{31}^2) =0.02\times 10^{-3}$eV$^2$.
The fact is that the total hadron energy is carried out by multiple low energy hadrons. 
The average energy per hadron is $\lapp$ 1 GeV and the average number of hadrons per
event is $\gapp$ 2.
The energy resolution of the hadrons at this energy is $\sim$ 80\%, and consequently, 
the neutrino energy resolutions do 
not improve significantly after adding the hadrons 
\footnote{The detection of  
the neutral current events may also be possible, which  have  no  
directional information, but the energy
dependence of the oscillation averaged over all directions can  contribute
to the total $\chi^2$ in the sensitivity studies.}.

The $\chi^2$ is calculated according to the Poisson probability distribution    
with flat uncertainties of the oscillation parameters.
The term due to the contribution of
prior information of the oscillation parameters measured by other
experiments  is not added to $\chi^2$ to examine solely the performance
of ICAL.
The data have been binned in cells of equal size  in
the  $\log_{10} E$ - $L^{0.4}$ plane, where $L=2 R \cos\theta_Z$.
The choice of binning is motivated by
pattern of the oscillation probability
$P(\nu_\mu \rightarrow \nu_\mu)$ in the
$L-E$ plane \cite{Samanta:2008ag}. The distance between two consecutive
oscillation peaks driven by $\Delta m^2_{23}$
increases (decreases) as one goes to lower $L$ ($E$) values for a given $E$ ($L$).
The binning of $L$ has been optimized to get better sensitivity
to the oscillation parameters. 
To maintain $\chi^2/d.o.f\approx 1$ for Monte Carlo simulation study,
number of events should be  $>4$ per cell \cite{Samanta:2008af}
(as large number of cells at high energies have number of events less than 4
or even zero and they increase the $\chi^2/d.o.f$ substantially beyond 1).
If the number is less than 4 (which happens in the high energy bins),
we combine events from the nearest cells. 

The migration of the number of events from true neutrino energy and zenith angle cells  to muon 
energy and zenith angle cells is made using exact energy-angle correlated 2-dimensional 
resolution functions \cite{Samanta:2006sj}.
%
For each set of oscillation parameters, we integrate the oscillated
atmospheric neutrino flux folding with the cross section, the exposure time, the target mass, the efficiency
and the two dimensional energy-angle correlated exact resolution functions to obtain the predicted data
in each cell in $L-E$ plane
for the $\chi^2$ analysis.
We use the charge current cross section of Nuance-v3 \cite{Casper:2002sd} and the
Honda flux  in 3-dimensional scheme \cite{Honda:2006qj}.
The number of
bins and resolution functions have been optimized in
\cite{Samanta:2008af}.
Both theoretical (fit values) and experimental (true values) data for $\chi^2$ analysis have been generated in the 
same way by migrating number of events from neutrino to muon energy and zenith angle bins 
using the resolution functions \cite{Samanta:2009qw}.

The systematic uncertainties of the atmospheric neutrino flux
are crucial for determination  of the oscillation parameters.
We have  divided them into two categories:
(i) the overall flux normalization uncertainties which are independent of
the energy and zenith angle, and
(ii) the spectral tilt uncertainties which depend on $E$ and $\theta_{\rm z}$.

The flux with uncertainties included  can be written as
\bea
&&\Phi (E,\theta_Z)  =  
\Phi_0(E) \left[ 1 + \delta_E \log_{10} \frac{E}{E_0} \right] \nonumber\\
     && \times \left[ 1 + \delta_Z (|\cos\theta_Z|-0.5) \right]  \times \left [1+\delta_{f_N}\right] \nonumber
\label{e:uncer}
\eea
For $E < 1$ GeV we take the energy dependent uncertainty   $\delta_E=15\%$ and $E_0=1$ GeV
and for $E>10$  GeV, $\delta_E=5\%$ and $E_0=10$ GeV.
The overall  flux uncertainty as a function of  zenith angle
is parametrized by $\delta_Z$.
According to  \cite{Honda:2006qj} we use  $\delta_Z=4\%$, which leads to 2\%
vertical/horizontal flux uncertainty.
We take  the overall flux normalization uncertainty
$\delta_{f_N}=10\%$
and  the overall neutrino cross-section uncertainty
$\delta_{\sigma}=10\%$.

We evaluate the $\chi^2$ 
for both normal hierarchy (NH) and inverted hierarchy (IH)
with $\nu$s and $\bar\nu$s separately for a given set of oscillation parameters.
Then we find the total $\chi^2$ ($=\chi^2_\nu+\chi^2_{\bar\nu}$).

We have set the inputs of $|\Delta m_{32}^2|=2.5\times 10^{-3}$eV$^2$, and $\delta_{CP}=0$.
We marginalize $\chi^2$ over 
$\Delta m_{32}^2,~\theta_{23}, ~\theta_{13}$ and $\alpha$.
We have chosen the range of $\Delta m_{32}^2=2.0-3.0\times 10^{-3}$eV$^2$,
$\theta_{23}=37^\circ - 54^\circ,$ $\theta_{13}=0^\circ - 12.5^\circ$,
and $\alpha_{e\mu,e\tau}=0-3\times 10^{-53}$. 
%
The solar parameters are fixed at their best-fit values:
$\Delta m_{21}^2 = 7.67 \times 10^{-5}$eV$^2$ and $\sin^2\theta_{12}=0.312$ \cite{Fogli:2008ig}.
The effect of $\Delta m_{21}^2$ comes in
sub-leading order in the oscillation probability for atmospheric neutrinos 
when $E\gapp$ GeV and it is very negligible.
%

\begin{figure*}[htb]
\includegraphics[width=7.0cm,angle=0]{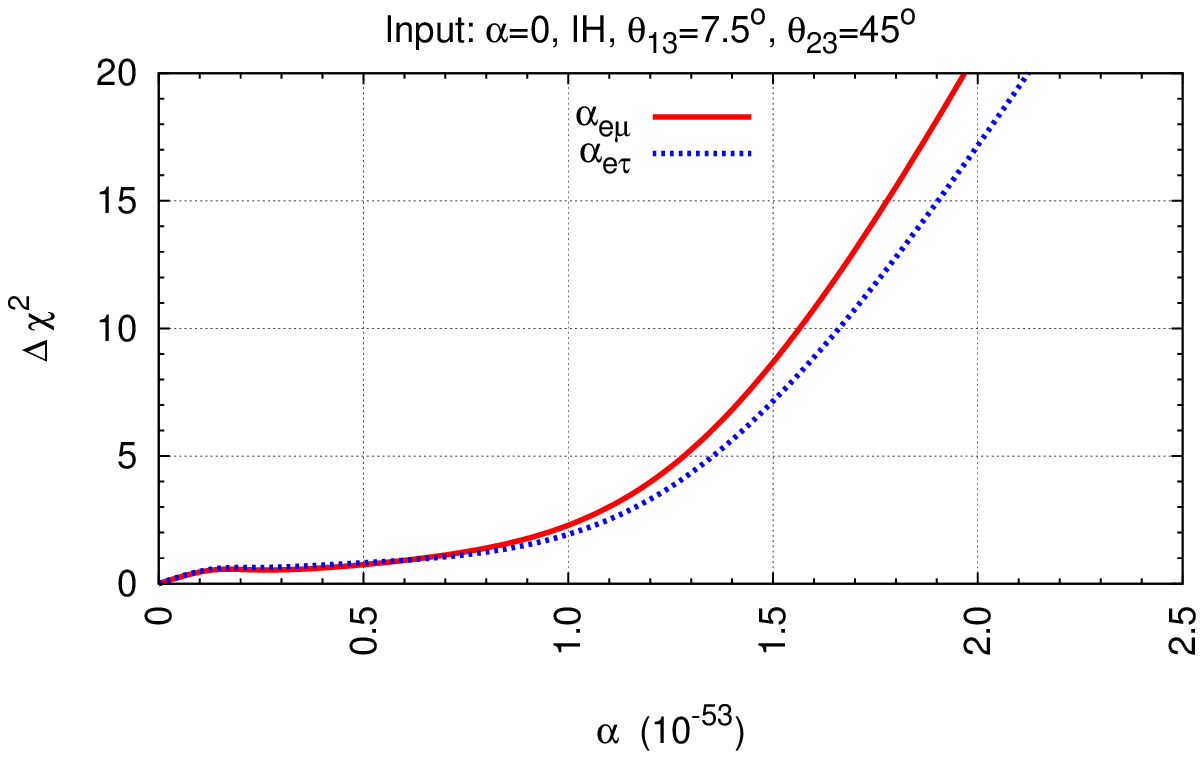}
\includegraphics[width=7.0cm,angle=0]{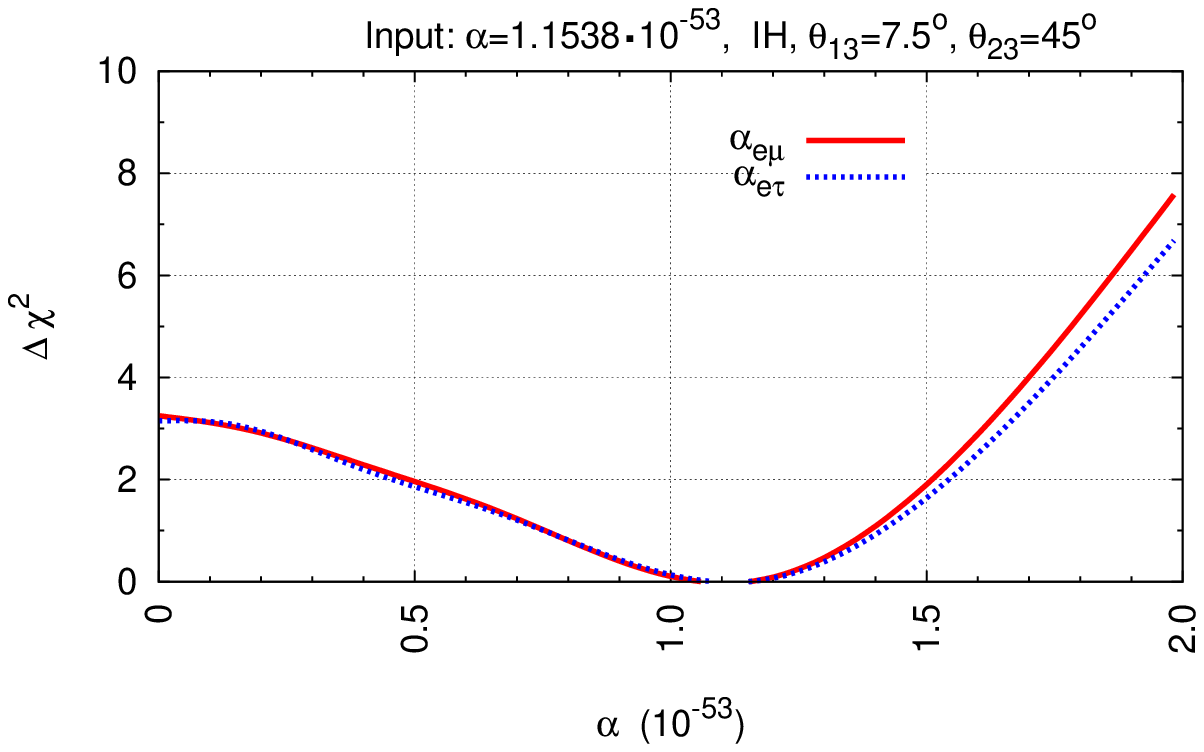}
\caption{\sf \small The  upper bounds of $\alpha_{e\mu}$ and  $\alpha_{e\tau}$
for  true (input) $\alpha=0$ in the left pannel; and the corresponding upper as well as lower bounds  for 
input $\alpha=1.1538\times 10^{-53}$ in the right pannel, 
respectively.   
}
\label{f:l0}
\end{figure*}
\begin{figure*}[htb]
\includegraphics[width=7.7cm,angle=0]{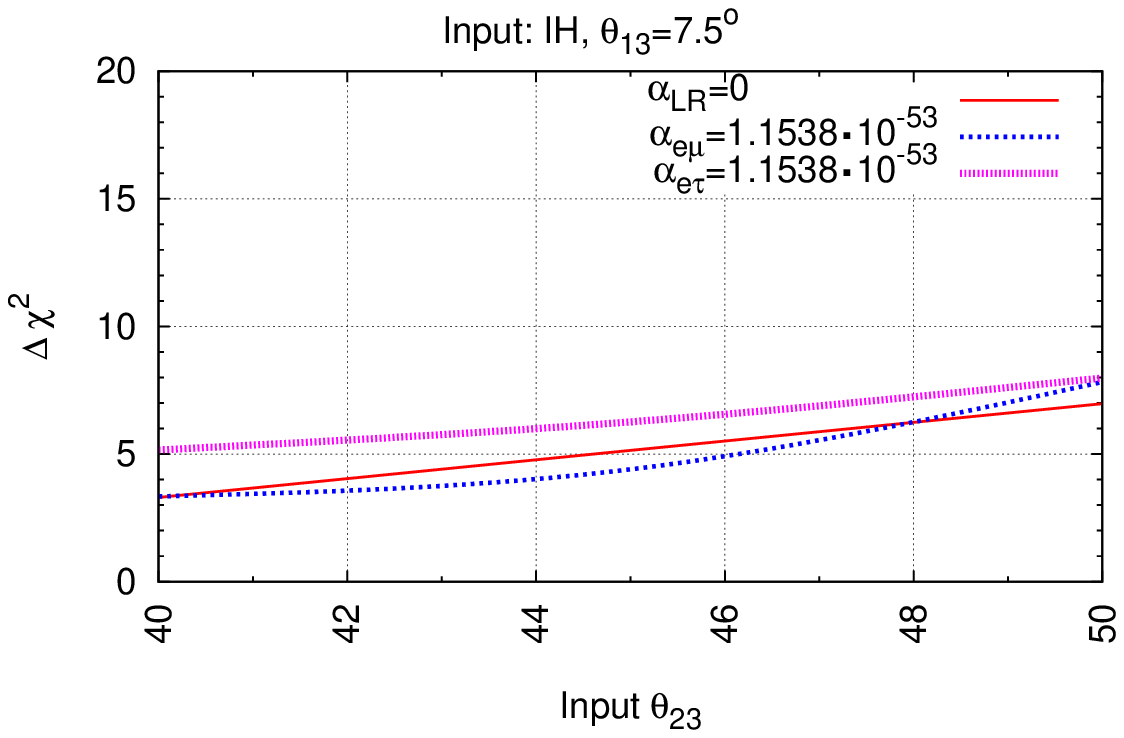}
\includegraphics[width=7.7cm,angle=0]{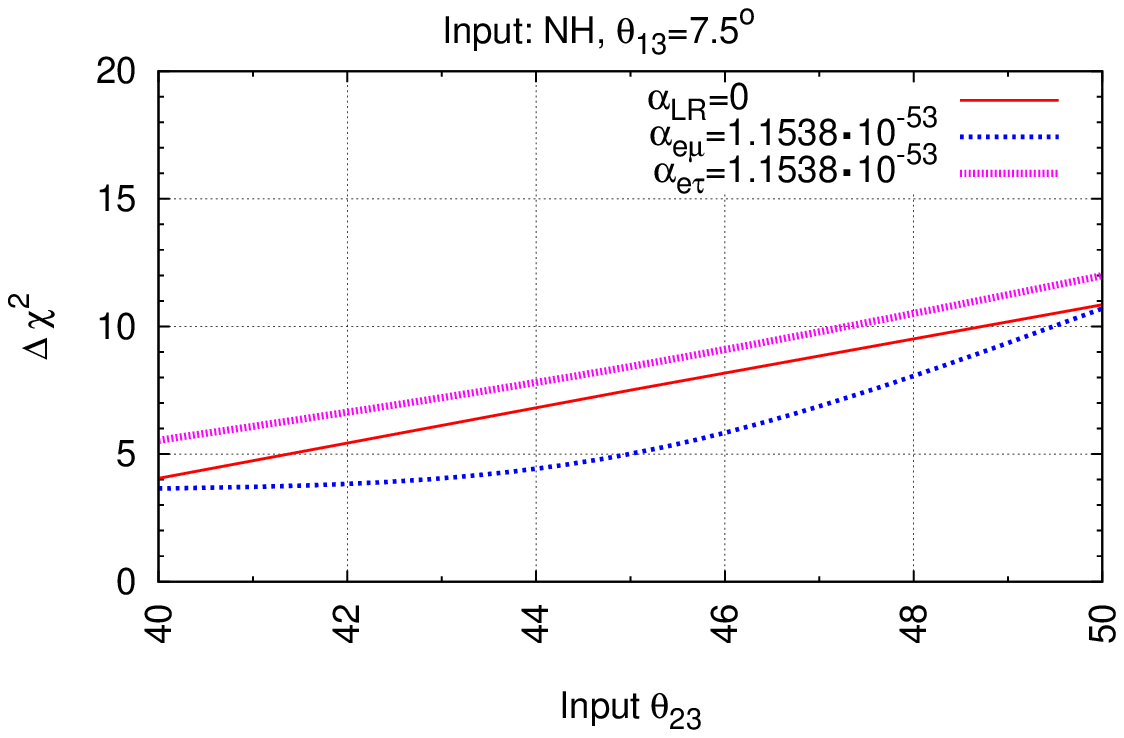}
\caption{\sf \small 
The  sensitivities to  mass hierarchy  for different true (input)
$\theta_{23}$ values for input  IH (left) and input  NH (right), respectively.
For each case we have considered true (input) values of $\alpha_{LR}=0$, $\alpha_{e\mu}
=1.1538\times 10^{-53}$, and $\alpha_{e\tau}=1.1538\times 10^{-53}$, respectively. 
}
\label{f:hier}
\end{figure*}

\begin{figure*}[htb]
\includegraphics[width=7.7cm,angle=0]{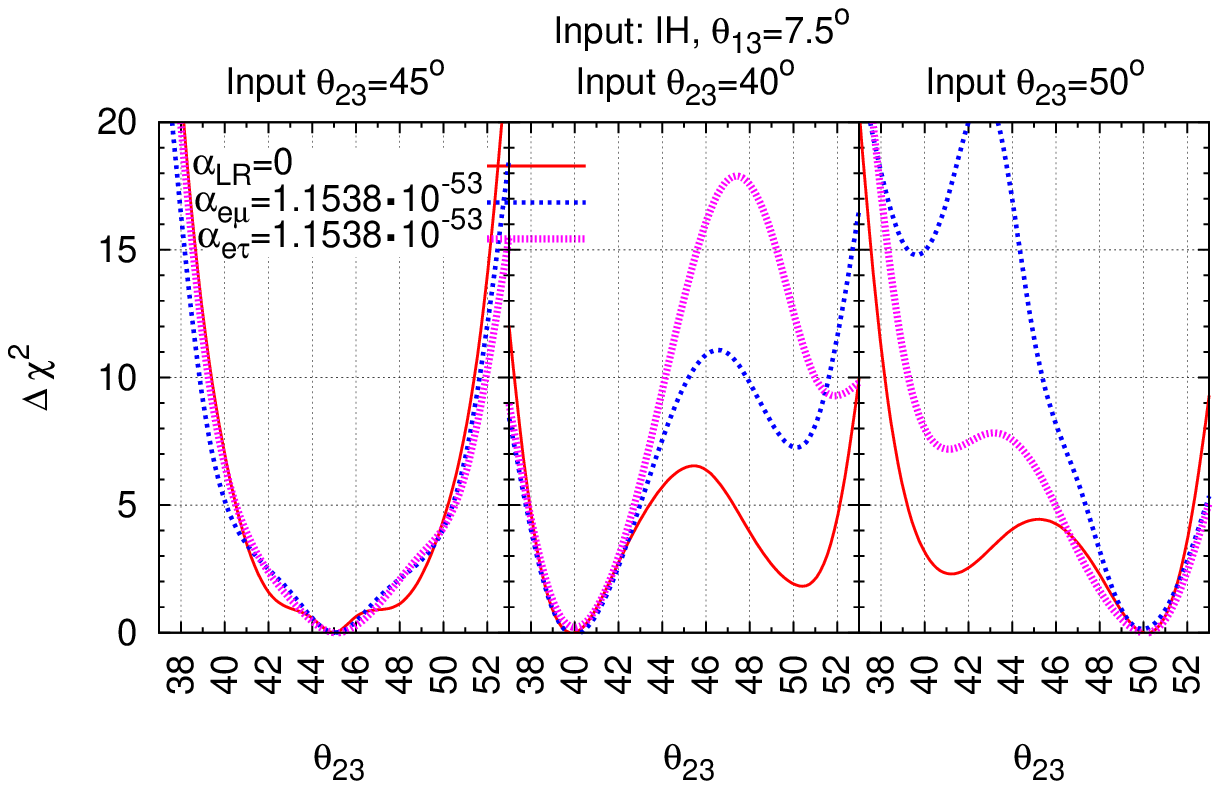}
\includegraphics[width=7.7cm,angle=0]{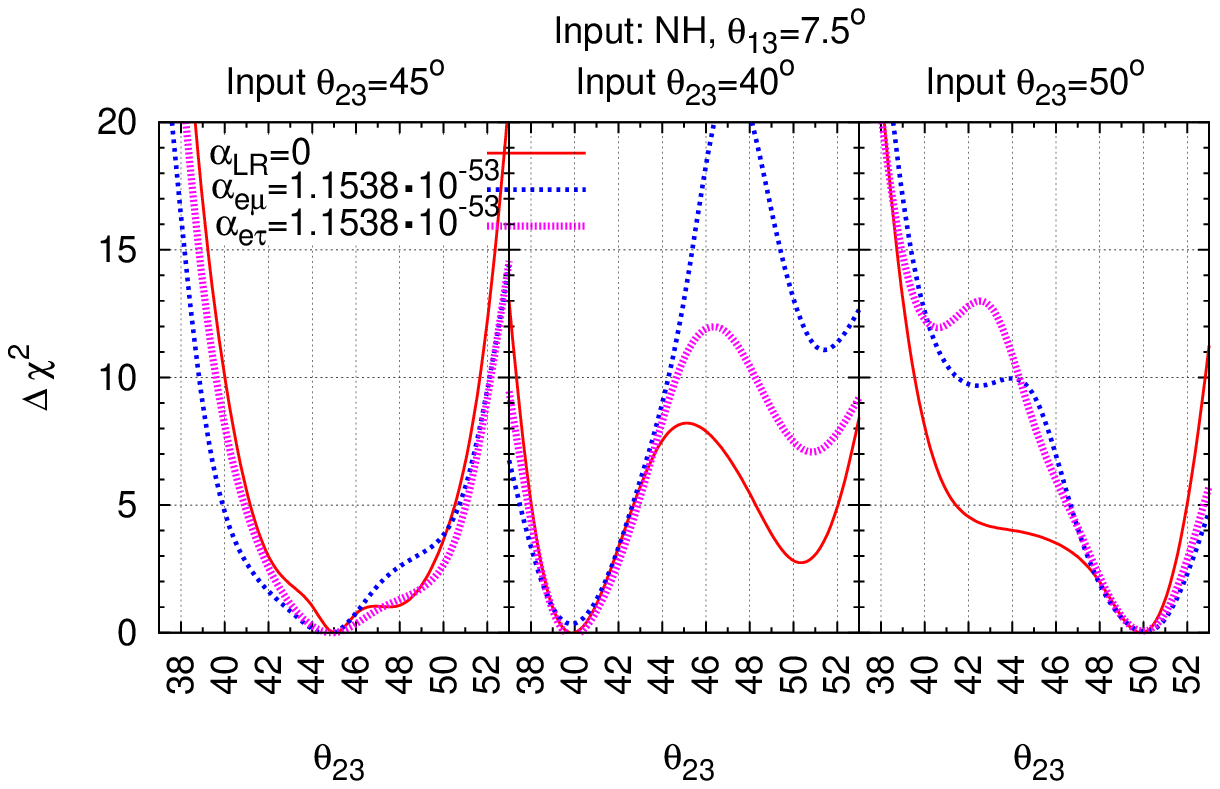}
\caption{\sf \small
The sensitivities to deviation of 2-3 mixing from maximal mixing as well as its octant
for both true (input)  IH (left) and NH (right). For each case we have considered  
$\alpha_{LR}=0$, 
$\alpha_{e\mu}=1.1538\times 10^{-53}$, $\alpha_{e\tau}=1.1538\times 10^{-53}$, respectively.
}
\label{f:lt23}
\end{figure*}

In this analysis we have considered an exposure of 1 Mton.year  
(which is 10 years run of 100 kTon) of ICAL and energy range 0.8 - 15 GeV.

\section{Bounds on couplings}\label{s:bounds}
For atmospheric neutrinos at the magnetized ICAL 
the bounds 
on  $\alpha_{e\mu}$ and $\alpha_{e\tau}$
 are shown in the left pannel in Fig. \ref{f:l0}  and its strength
 to constrain them with upper as well as  lower bounds 
are shown in the right pannel in Fig. \ref{f:l0} with a typical nonzero  
input (true) value.
We have checked that 
the bounds do not change significantly
for  different combinations 
of true values of oscillation parameters.


The bounds  $\alpha_{e\mu} \lapp 5.5 \times 10^{-52}$ and $\alpha_{e\tau} \lapp 6.4 
\times 10^{-52}$ at 90\% CL  have been obtained from  present atmospheric neutrino 
oscillation data \cite{Joshipura:2003jh}.
Considering both atmospheric and solar data the 
bounds are $\alpha_{e\mu} \lapp 3.4 \times 10^{-53}$  and $\alpha_{e\tau} \lapp 2.5 \times 10^{-53}$ 
at $3\sigma$ CL  \cite{Bandyopadhyay:2006uh}. 
The  better precision and smaller value by a few order of magnitude of solar mass squared difference 
 than the atmospheric one play here the main role to make it significantly stringent. 

The bounds are relatively stronger at the magnetized detector: 
$\alpha_{e\mu,e\tau} \lapp 1.65\times 10^{-53}$ at 3$\sigma$ CL. 
For atmospheric neutrinos it comes mainly due to
the high precision of $\Delta m_{32}^2$ and this can be understood quite well
from simple two flavor survival probability of $\nu_\mu$ and $\bar\nu_\mu$
in vacuum with effective $\Delta m_{32}^2$ and $\theta_{23}$ for long-range potentials. 
This becomes possible due to the fact that the  1-3 mixing effect  
is sub-leading (as $\theta_{13}$ $<11.38^\circ$ \cite{Schwetz:2011qt})
and it can be neglected at this moment for simplicity. 
When the potential $V_{LR}$ comes to the play, it tries to change the 
effective value of 
 $\Delta m_{31}^2$ (see Eq. \ref{e:dM})  and becomes tightly constrained.
The sign  $V_{LR}$ is opposite for $\nu$ and $\bar\nu$, and ICAL
can detect them separately. This makes the bounds more tighter at ICAL
and lessens
the difference 
in bounds between $\alpha_{e\mu}$ and $\alpha_{e\tau}$, while the difference is 
substantially large
 at non-magnetized detectors.


\section{Effects on  2-3 sector}\label{s:23sector}

\subsection{Determination of mass hierarchy} In Fig. \ref{f:hier} we show the sensitivity to 
 mass hierarchy for different true values of $\theta_{23}$. 
We show it for three cases:  assuming no potential for long-range forces 
($\alpha_{LR}=0$) and with a benchmark input (true) value for  
$\alpha_{e\mu}$ and $\alpha_{e\tau}$, respectively.

For all three cases 
the sensitivity increases as one goes to the higher $\theta_{23}$ values and it is 
a general feature for both cases with and without the potential for 
long-range force. 
This can be understood from the  $\nu_\mu$ flux 
at the detector considering the effect of 1-3 mixing.
Here we assume no potential for long-range force. 
The ratio of the $\nu_\mu$ flux at the detector (oscillated) 
and at the source (original) \cite{Samanta:2010xm}:
\be
\frac{ F_\mu}{F_\mu^0} \approx
K(\sin 2 \theta_{23}) -
f (\theta_{23}) \left( 1 - \frac{1}{r}\right) P_A (\theta_{13}),
\label{eq:flap}
\ee
where,  $K(\sin 2 \theta_{23}) $ is an even function of the
 deviation (symmetric with respect to
change of the octant), and  $P_A (\theta_{13})$ is a function of
$\theta_{13}$ only, and
\be
f (\theta_{23})  \equiv  \left(s_{23}^4 - \frac{s^2_{23}}{r} \right)
\ee
which increases quickly  with  $\theta_{23}$,  so that
for $r =  3 - 4$,     $f (\theta_{23} = 40^{\circ}) \ll
f (\theta_{23} > 50^{\circ})$. 
Therefore,
for  $\theta_{23} < 45^{\circ}$ the flux $F_\mu$
has much weaker  dependence on $\theta_{13}$ than for
$\theta_{23} > 45^{\circ}$.
This is reflected in the 
sensitivity to  mass hierarchy as the lower limit of $\theta_{13}$ 
has been taken zero during marginalization 
and no prior contribution from other future experiments has been considered. 
However, in future
the lower limit will be known from other experiments 
(as very recently  T2K puts a lower bound \cite{Abe:2011sj});
then the $\theta_{23}$-dependence will be less and the sensitivity to 
 mass hierarchy will also be substantially improved \cite{Samanta:2009qw}.  
 
The potential  $V_{e\mu}$ opposes,  while  $V_{e\tau}$ 
helps  to determine the hierarchy (see Fig. \ref{f:hier}). 
As the 1-3 mixing is small ($\theta_{13} < 11.2^\circ$),  the potential $V_{e\tau}$
decreases the effective value of $\Delta m_{31}^2$. 
Consequently, the resonant energy decreases (see Fig. 2 of \cite{Samanta:2006sj}),
where the atmospheric neutrino flux is relatively large and it helps in determination of 
hierarchy. On the other hand, 
as 2-3 mixing is large,  $V_{e\mu}$ increases the effective value of  $\Delta m_{31}^2$ 
for $\theta_{23}>45^\circ$ (see Eq. \ref{e:dM}).  
In spite of an enhancement in  1-3 mixing (which happens for both potentials)
the final sensitivity decreases  for $V_{e\mu}$ due to relatively low statistics 
at the resonant zones. These happen for $\nu$ with NH and for 
$\bar\nu$ with IH.
 

\subsection{Determination of octant of $\theta_{23}$}
The sensitivity to 2-3 mixing, mainly the deviation from its maximal mixing ($\delta=\theta_{23}-45^\circ$)
 and the 
octant (sign of $\delta$) are shown in Fig. \ref{f:lt23} for three cases:  i) $\alpha_{LR}=0$,
ii) $\alpha_{e\mu}=1.1538\times 10^{-53}$, and iii) $\alpha_{e\tau}=1.1538\times 10^{-53}$, 
respectively.
It is important to note here that octant discrimination  is significantly improved 
in presence 
of  long-range potentials. This is due to the fact that effective 1-3 mixing is always larger
for both potentials. For a given hierarchy the change of the sensitivity 
with the change of true (input) octant
depends mainly on the magnitude of  $\delta^{\rm eff}={\theta_{23}}^{\rm eff}
-45^\circ$, which again strongly depends on the potentials 
(see Eq. \ref{e:sin}).
The  change  of the effective value of $\Delta m_{31}^2$ works here subdominantly (while it was a dominating factor for determination of the mass hierarchy).
Now, for $V_{e\tau}$ with NH, sensitivity to octant determination 
 is better for $\theta_{23} > 45^\circ$ than $\theta_{23} < 45^\circ$ 
as $\delta^{\rm eff}$ is increased
due to $V_{e\tau}$ for neutrinos and 
the flux  is two times higher for neutrinos than antineutrinos. 
This is opposite
 for $V_{e\mu}$:  octant determination
 is better for $\theta_{23} < 45^\circ$ than $\theta_{23} > 45^\circ$.
Similarly, the results with IH can be understood using antineutrinos
considering the symmetries in Eq. \ref{e:ppba}.

\section{MINOS anomaly}\label{s:minos}
The anomaly in recent MINOS data  (difference in measured values of 2-3 
mass splittings and mixing angles for $\nu$ and $\bar\nu$ \cite{pvahle}), 
which implies CPT violation
or signal for non-standard interactions,  has been explained by 
long-range potential due to $L_\mu - L_\tau$ gauge symmetry \cite{Heeck:2010pg}. 
The Sun does not contain any $\mu$ or $\tau$ and hence there is no direct bound from 
neutrino oscillation.  However, the potential can be induced indirectly on the Earth by 
the mixing of the gauge boson $Z^\prime$ of $L_\mu-L_\tau$  
with the standard model $Z$ boson \cite{Heeck:2010pg}:
\be
V_{\mu\tau}= 3.6\times 10^{-14}{\rm eV} \left(\frac{\alpha_{\mu\tau}}{10^{-50}}\right)
\ee
The sign changes for $\bar\nu$. 
From the  analysis of atmospheric neutrino data at ICAL with an exposure of 1 Mton$\cdot$yr,
we find the bound on the coupling $\alpha_{\mu\tau} \ge 3.2 (4.2) \times 10^{-51}$ at 3(5)$\sigma$ 
CL and the explanation of MINOS data \cite{Heeck:2010pg} can be tested
at more than 5$\sigma$ CL.

\section{Discussion and Conclusion}\label{s:dis}
In this paper, we have estimated the conservative bounds on long-range forces 
with atmospheric neutrinos at ICAL:
$ \alpha_{e\mu, e\tau} \lapp 1.65 \times
10^{-53}$ at 3$\sigma$ CL. This bounds are significantly 
stronger than the present bounds. 

The CPT violation and the long-range forces, which are the candidates for  solution of  
recent  anomaly in
MINOS data, can be discriminated with atmospheric neutrinos at ICAL. 
The distance between the Sun and the Earth varies and the difference between 
aphelion and perihelion  is about 3\%, which causes seasonal variation of the 
long-range forces. But, there should not be any such change 
for CPT violation.

Assuming one by one nonzero input value of these couplings
it is found that while the potential $V_{e\mu}$ opposes; 
$V_{e\tau}$ helps to discriminate the mass hierarchy.
However, both potentials help to discriminate octant of $\theta_{23}$.
The anomaly of MINOS data, which has been explained using long-range force potential,  
can be tested at ICAL with more that 5$\sigma$ CL.

{\it Acknowledgements:}
This work was started at Harish-Chandra Research Institute and has been supported partly by the Neutrino Physics projects
of this institute. 
The  use of cluster computational
facility installed by the funds of this project is gratefully acknowledged.

\end{document}